\documentclass[aps,superscriptaddress,amsmath,amssymb,showpacs,twocolumn,prb]{revtex4}
\usepackage{hyperref}
\usepackage{epsfig}
\usepackage{subfigure}
\usepackage{braket}
\usepackage{bm}
\usepackage{color}
\usepackage{float}
\usepackage[babel=true]{csquotes}

\newcommand{\rem}[1]{}

\newcommand{\beq}{\begin{equation}}
\newcommand{\eeq}{\end{equation}}
\newcommand{\beqa}{\begin{eqnarray}}
\newcommand{\eeqa}{\end{eqnarray}}

\newcommand{\refe}[1]{\eqref{#1}}
\newcommand{\refE}[1]{Eq.~\eqref{#1}}

\newcommand{\Tr}{\mathrm{Tr}}
\newcommand{\re}{\mathrm{Re}}
\newcommand{\im}{\mathrm{Im}}
\newcommand{\cht}{\cosh{\vartheta}}
\newcommand{\sht}{\sinh{\vartheta}}

\newcommand{\eq}{&=&}

\begin{document}
\title{Signatures of odd-frequency correlations in the Josephson current of superconductor/ferromagnet hybrid junctions}

\author{Caroline Richard} 
\affiliation{Univ.~Grenoble Alpes, INAC-SPSMS, F-38000 Grenoble, France, and
\\CEA, INAC-SPSMS, F-38000 Grenoble, France}
\author{Alexandre Buzdin}
\affiliation{University Bordeaux, LOMA UMR-CNRS 5798, F-33405 Talence Cedex, France}
\author{Manuel Houzet}
\affiliation{Univ.~Grenoble Alpes, INAC-SPSMS, F-38000 Grenoble, France, and
\\CEA, INAC-SPSMS, F-38000 Grenoble, France}
\author{Julia S. Meyer}
\affiliation{Univ.~Grenoble Alpes, INAC-SPSMS, F-38000 Grenoble, France, and
\\CEA, INAC-SPSMS, F-38000 Grenoble, France}

\begin{abstract}
Contacting a bilayer ferromagnet with a singlet even-frequency superconductor allows for the realization of  an effective triplet odd-frequency superconductor. In this work, we investigate the Josephson effect between superconductors with different symmetries (e.g. odd- versus even-frequency).  {  In particular, we study the  supercurrent flowing between two triplet odd-frequency superconducting leads through a weak singlet  even-frequency superconductor. We show that  the peculiar temperature dependence of the critical current below the superconducting transition of the weak superconductor is a signature of the competition between  odd/odd-frequency  and odd/even-frequency Josephson couplings.} 

\end{abstract}
\date{\today}

\pacs{74.45.+c, 74.50.+r, 75.70Cn, 74.20.Rp}

\maketitle


\section{introduction}

It is well known that superconductivity arises from the formation of Cooper pairs of electrons, where the wavefunction of a pair is a function of spin, space and frequency (or time). The Pauli principle tells us that such a wavefunction  should be antisymmetric. In dirty metals, due to multiple scattering events on impurities, the orbital part is necessarily symmetric. As a consequence, a spin-singlet pairing  is even  in frequency and a spin-triplet pairing is odd in frequency.~\cite{eschrig}
In conventional superconductors ($S$) the pairing is even in frequency. However,  it has been predicted  that, thanks to the proximity effect, one may induce  spin-triplet odd-frequency correlations in hybrid superconducting/ferromagnetic structures ($S/F$).  When $F$ is homogeneous, the triplet proximity effect involves electrons of opposite spins and is short-ranged.~\cite{buzdin-rev} By contrast, an inhomogeneous magnetization, as, e.g., in  non-collinear bilayer ferromagnets ($F'/F$), also induces long-range triplet correlations between electrons with parallel spins.~{\cite{bergeret,bergeret-rev}} Furthermore, if $F'$ is  short  whereas $F$ is much longer than the coherence length of singlet correlations,  \enquote{pure} triplet odd-frequency correlations are induced at the extremity of the long ferromagnet. Thus the $S/F'/F$ structure realizes an  effective triplet odd-frequency reservoir ($S_T$).  In this paper, we study how to probe these odd-frequency correlations.

Recently long-range supercurrents have been measured  in trilayer ferromagnetic Josephson junctions, which  can be viewed as Josephson junctions between two odd-frequency reservoirs ($S_T/S_T$ junctions).~\cite{klose, robinson, birge} While this  indicates the presence of triplet odd-frequency correlations, the  measurements  did not present any peculiarities as compared to observations made in \enquote{classic} Josephson junctions connecting two conventional superconductors ($S/S$).  Indeed, from a symmetry point of view   the $S_T/S_T$ junction as well as the $S/S$ junction  realize  a  coupling  between two reservoirs sharing the same symmetry (odd/odd-frequency for $S_T/S_T$ and  even/even-frequency for $S/S$), yielding similar supercurrent measurements. 
By contrast, the current-phase relation of $S_T/S$ junctions is predicted to be superharmonic.~\cite{radovic,trifunovic, superh} This specificity originates from the odd/even-frequency Josephson coupling. Namely,  the symmetry mismatch  between the  reservoirs prohibits  mechanisms involving the transfer of a single Cooper pair. Instead, the supercurrent originates from the coherent flow of an even number of pairs, yielding a peculiar $\pi$-periodic current-phase relation. 

\begin{figure}[h]
(a)
\includegraphics[width=0.9\linewidth]{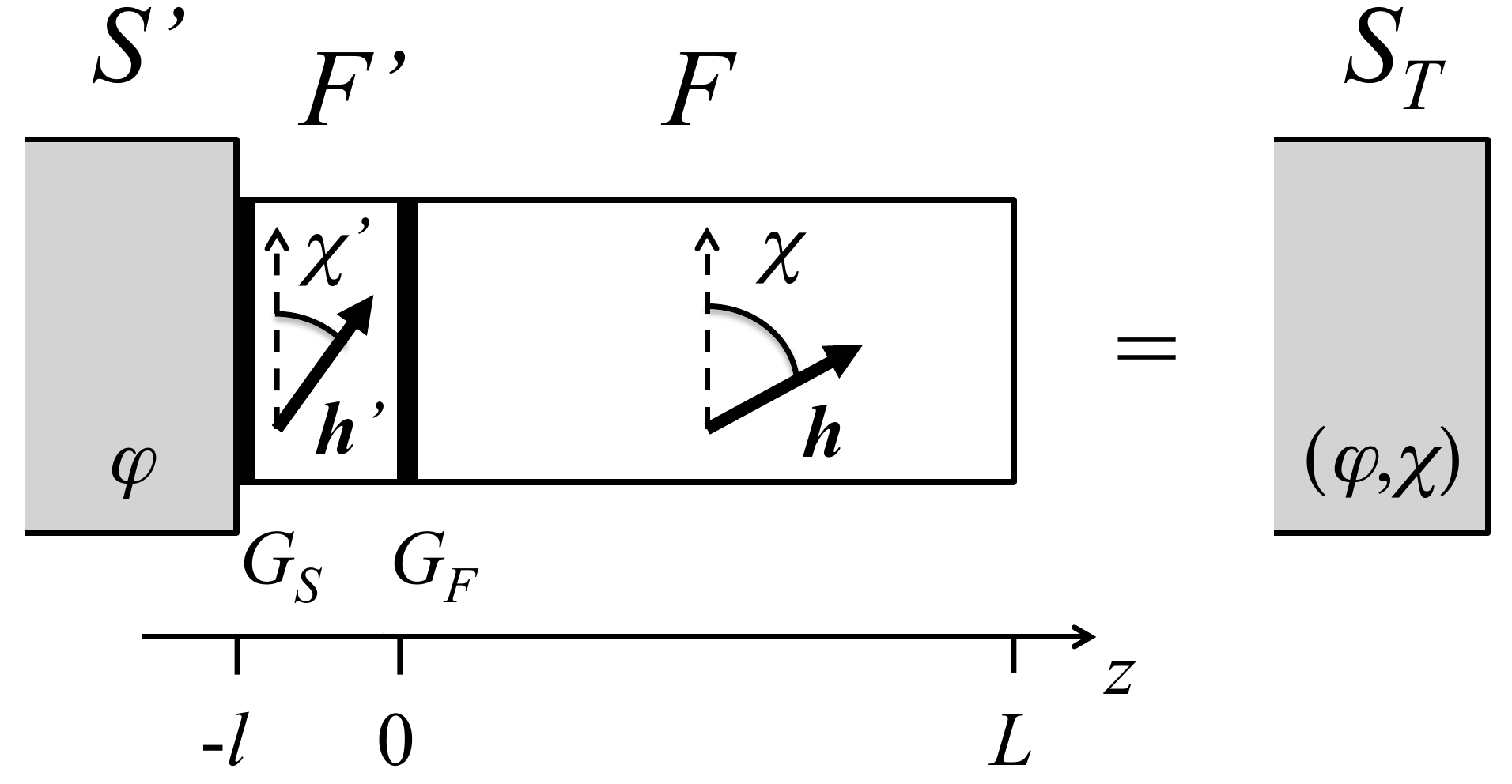}
\\
(b)
\includegraphics[width=0.6\linewidth]{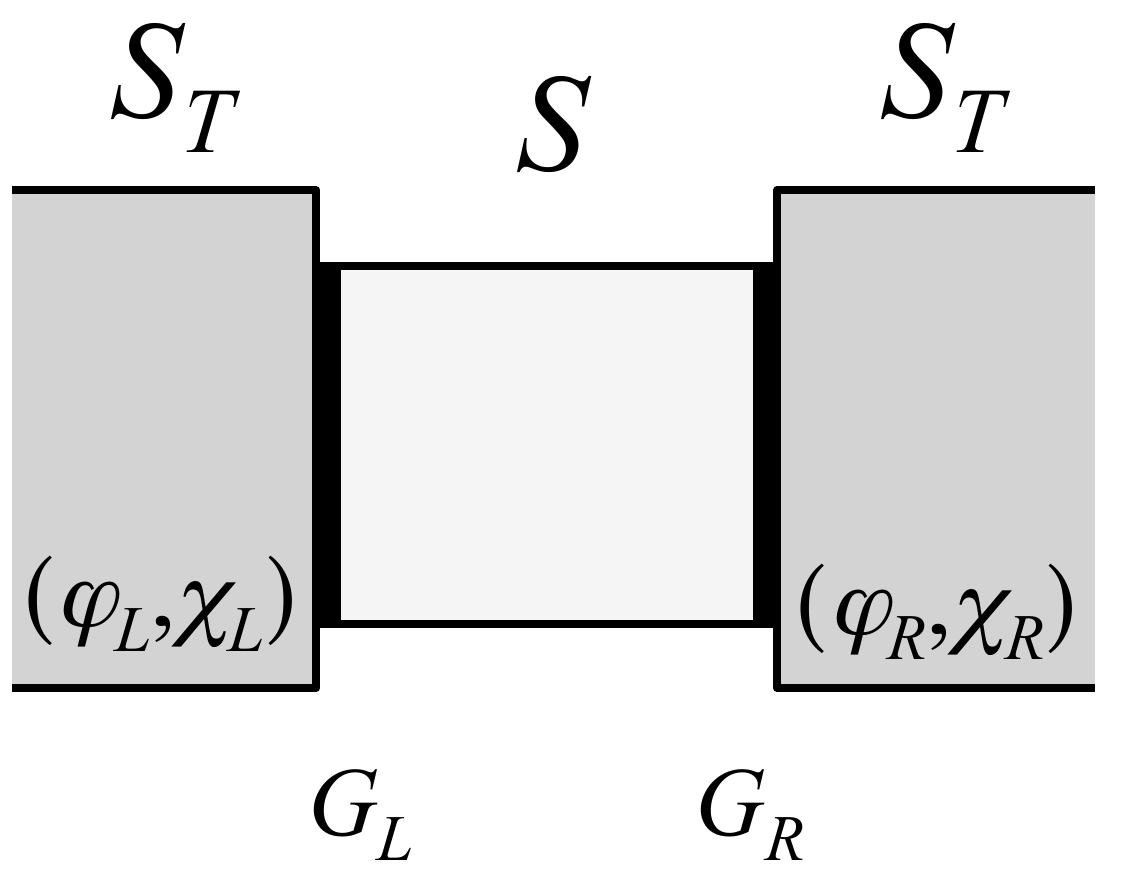}
\caption{{ 
(a) Realization of an effective triplet odd-frequency superconductor $S_T$ by contacting a ferromagnetic $F'/F$ bilayer with noncolinear magnetizations, with a relative angle $\theta=\chi-\chi'$ between their directions, to a singlet even-frequency superconductor $S'$. 
(b) Effective $S_T/S/S_T$ double Josephson junction formed with triplet odd-frequency superconducting leads connected through a singlet even-frequency superconducting layer. 
}}
\label{Fig:setup}
\end{figure}

In this work, we explore  the  competition  between  odd/odd-frequency  and odd/even-frequency Josephson couplings through the temperature dependence of  the critical current of hybrid junctions. In particular,  we   study the current through an $S_T/S/S_T$ junction, where a  conventional  superconductor of bare critical temperature $T_c$  is sandwiched between two effective triplet odd-frequency reservoirs. Such a junction may be realized in a $S'/F'/F/S/F/F'/S'$ hybrid junction, where    $S'$ are  conventional superconductors  with a critical temperature $ T_c'$,
{ see Fig.~\ref{Fig:setup}.} 


In the following, we assume that $T_c'\gg T_c$.
Above the critical temperature $ T_c$  of the superconducting layer, an effective odd/odd-frequency  Josephson coupling builds up: the transfer of \enquote{odd-frequency} pairs between leads happens via virtual Andreev pairs in the island. Below $T_c$,  when the central $S$-layer  is superconducting,   the quasi-particles  above the gap  coexist with the even-frequency condensate of Cooper pairs. Therefore, an additional odd/even-frequency Josephson coupling is generated at the interfaces between the layer and the leads, generating a double  $S_T/S$ Josephson junction.

We will show that the currents associated with odd/odd-frequency and odd/even-frequency Josephson couplings  are in competition. Besides a  peculiar current-phase relation,  {  this leads to a suppression of the critical current below the transition temperature of the weak superconductor.} 
 
The outline of the paper is as follows: in Sec.~\ref{sec-formalism}, we introduce the formalism and, in Sec.~\ref{sec-reservoirs}, we derive the Green function of an effective  $S_T$ reservoir. Then, in Sec.~\ref{sec-junction}, we treat the full $S_T/S/S_T$ junction and compute both its current-phase relation and its critical current. Finally, we briefly discuss metallic junctions in Sec.~\ref{sec-metallic}, before concluding in Sec.~\ref{sec-conclusion}.

\section{formalism}
\label{sec-formalism}

Within the quasi-classical theory, the equilibrium properties of hybrid superconducting/ferromagnetic junctions can be expressed via the quasi-classical Matsubara Green function $g$, which is a $4\times4$ matrix in the particle-hole and spin spaces, and obeys the normalization conditions $g^2=1$ and $\Tr[g]=0$. Within circuit theory,~\cite{circuit-theory} $g$ takes the value $g_i$ in each superconducting or ferromagnetic node or reservoir of the circuit. The Green functions in the nodes obey the equations
\begin{equation}
\label{dot}
\frac {2\pi G_Q }{\delta_i }[(\omega+  \bm{h}_i.\bm{\sigma}) \tau_z +\hat \Delta_i, g_i ] +\sum_j\hat I_{ij}=0.
\end{equation}
Here, $\delta_i$ is the mean level spacing in  node $i$, $G_Q=e^2/\pi$ is the conductance quantum, $\omega= (2n+1)\pi T $ is a positive Matsubara frequency at temperature $T$ ($n\geq 0$), $\bm{h}_i$ is an exchange field acting on the electron spin in ferromagnetic nodes, and $\hat \Delta_i=\Delta_i(\cos\varphi_i\tau_x-\sin\varphi_i\tau_y)$, where $\Delta_i$ and $\varphi_i$ are the modulus and phase of the superconducting order parameter in superconducting nodes. The Pauli matrices $\tau_i$ and $\sigma_i$ ($i=x,y,z$) act in particle-hole and spin space, respectively. Moreover, the spectral current between two nodes or leads $i$ and $j$,
\begin{equation}
\hat I_{ij}= \frac {G_{ij}}2[g_j, g_i],
\end{equation}
is related with the normal-state conductance $G_{ij}$ of the \enquote{connector} between them. The current flowing through that connector is
\begin{equation}
\label{currentf}
I_{ij}=-\frac {\pi  T}{2e}{\rm Im} \sum_{\omega>0} 
\Tr [ \tau_z\hat I_{ij}].
\end{equation}
Finally, in the case of a superconducting node with \enquote{bare} critical temperature $T_{ci}$, the order parameter should satisfy the self-consistency equation
\begin{equation}
\label{self}
\ln \frac  {T_{ci}} T=2\pi T \sum_{\omega>0}
\left(
\frac1\omega-
\frac{e^{-i\varphi_i}}{4\Delta_i}
 \Tr\left[\tau_-( {g + g^\dagger})\right]
\right),
\end{equation}
where $\tau_-=(\tau_x-i\tau_y)/2$.

Note that  Eqs.~\refe{dot} and  \refe{self} account for the current  conservation  in each node $i$, namely $\sum_{j} I_{ij}=0$ is automatically satisfied.

\section{triplet odd-frequency reservoirs}
\label{sec-reservoirs}

As mentioned in the Introduction, a trilayer structure consisting of a conventional superconductor and two non-collinear ferromagnets realizes an effective triplet reservoir ($S_T\equiv S'/F'/F$), { see Fig. \ref{Fig:setup}(a).} In this section we derive the Green function of $S_T$.
We take the $z$-axis perpendicular to the layers and choose the magnetizations of both ferromagnetic layers to lie in the $xy$-plane, namely
\begin{equation}
\begin{cases}
\bm h'= h'(\cos\chi'\,\hat x + \sin\chi' \,\hat y )& \mbox{ in  } F',\\
\bm h= h(\cos\chi\,\hat   x + \sin\chi\, \hat y )& \mbox{ in }F.
\end{cases}
\end{equation}
Hence $\xi_F'=\sqrt{D/h'} $ and  $\xi_F=\sqrt{D/h} $ are the ferromagnetic coherence lengths in $F'$ and $F$, respectively.

The first layer $F'$ generates only short-range correlations. Thus, its length $l$ should not exceed the ferromagnetic coherence length $\xi_F'$. By contrast, the non-collinear second layer $F$ generates triplet correlations with all different spin projections. To filter out only the long-range components, its length $L$ needs to be much longer than $\xi_F$.

Within the quasi-classical theory, we call $g_S$, $g_{F'}$, and $g_F$ the Green function in the $S'$,  $F'$, and $F$ layers, respectively. Here, $S'$ is a reservoir. Thus, in the subgap regime, $\omega\ll T_c'$,  the Green function $g_S$ takes the form $g_S=\cos\varphi\,\tau_x-\sin\varphi\,\tau_y\equiv\tau_\varphi$, where $\varphi$ is the superconducting phase. The Green functions in the ferromagnetic layers will be determined in the following, using the formalism introduced in Sec. \ref{sec-formalism}. In particular, the Green function $g_T$ of our effective triplet reservoir is   related to correlations developing at the edge of $F$, namely $g_T=g_F(L)$.

Assuming $l\ll \xi_F'$, we can use circuit theory, where the $F'$ layer is  a ferromagnetic node. Its Green function $g_{F'}$  obeys
\begin{equation}
[\frac {2\pi G_Q }{\delta_{F'}}(\omega+ ih'\sigma_{\chi'}) \tau_z +\frac12(G_S g_S+G_F g_F), g_{F'}] =0,
\end{equation}
where  $\delta_{F'}$ is the mean level spacing in $F'$ and $G_{S}$  ($G_F$) is  the conductance of  the $S'/F'$ ($F'/F$) interface. Furthermore, we introduced the short-hand notation $\sigma_{\chi'}=\cos\chi'\,\sigma_x+\sin\chi'\,\sigma_y$. 
In the following, we assume that  $F'$ is more strongly coupled to $S'$, i.e.,  $G_S\gg G_F$, and neglect the leakage current at the $F'/F$ interface to obtain
 \begin{equation}
 \label{resSF}
 g_{F'}=\frac {(\omega+i  h'\sigma_{\chi'})\tau_z+\gamma_S\tau_\varphi}{\sqrt{(\omega+ i h'\sigma_{\chi'})^2+\gamma_S^2}}.
 \end{equation} 
Here $ \gamma_S= \delta_{F'} G_S/( 2\pi G_Q)  $ is the induced minigap in $F'$. Note that the same Green function with $\gamma_S=\Delta>h'=E_Z$ describes a superconductor subject to an external Zeeman field $E_Z$. The advantage of using an $S'/F'$ bilayer is  the possibility  of realizing both regimes $h'<\gamma_S$ and $h'>\gamma_S$ by tuning, e.g.,  the transparency  of the $S'/F'$ interface ($G_S$) or the thickness of the $F'$  layer ($ \delta_{F'}\propto 1/ l$). 

Having determined $g_{F'}$, we now turn to the long ferromagnetic layer $F$ of length $L\gg\xi_F$.   Close to the $F'/F$ interface, both short- and long-range correlations coexist. The fast oscillatory behavior of the short-range correlations prevents us from directly applying the circuit theory. However,  within a few $\xi_F$ from the $F'/F$ interface, the short-range correlations are suppressed, and only the non-oscillating long-range triplet correlations survive. Then, for $\xi_F \ll z \ll\xi_N$ with $\xi_N=\sqrt{D/2\pi T}$, we find  $ g_F(z)\approx g_T= \rm const$. In particular,  if $\xi_F\rightarrow 0$, the Green function may be considered constant throughout the layer. Thus, within a circuit theory approach, the long $F$ layer  maps to a ferromagnetic node with $\xi_F\rightarrow 0$ or, correspondingly, $h\rightarrow \infty$. Consequently $g_T$ obeys the equation,
\begin{equation}
\label{infin}
[\omega \tau_z + i  h\sigma_\chi\tau_z+\gamma_F g_{F'}, g_T]=0\enspace \mbox {with }\, h\rightarrow \infty,
\end{equation}  
where $\gamma_F=\delta_F G_F/ ( 2\pi G_Q) $, and  $\delta_F$ is the mean level spacing in $F$. 

To determine $g_T$, we orthogonally decompose the Green function $g_{F'}$ with respect to the length scale over which it decays  in $F'$. Namely, we write $g_{F'}= g_\parallel +g_\perp$, where $[g_\parallel, \sigma_\chi\tau_z]=0$ and $\{g_\perp, \sigma_\chi\tau_z\}=0$. Here $g_\parallel$ contains the long-range correlations, whereas $g_\perp$ contains the short-range correlations in $F$.  Then, $g_\parallel$ may be further decomposed by noting that terms $\propto \sigma_\chi\tau_z$ can be absorbed into $h$ in Eq.~\eqref{infin}. Thus, we write $g_\parallel$ as $g_\parallel= \tilde g_\parallel + J\sigma_\chi\tau_z$ with   $J=   \frac 14\Tr[ \sigma_\chi \tau_zg_\parallel]$.

In the limit $h\rightarrow \infty$, the short-range correlations are completely suppressed, while the long-range correlations are not affected by $h$. As shown in appendix \ref{app-hinf}, \refE{infin} may   be rewritten in the form
\begin{equation} 
[\omega \tau_z +\gamma_F \tilde g_\parallel, g_T]=0,
\label{dotF}
\end{equation}
where $\tilde g_\parallel(\omega)= \alpha(\omega)\tau_z+i\beta (\omega)\sigma_\chi\tau_ \varphi$ can  be obtained from Eq.~\eqref{resSF} with
\begin{eqnarray}
\alpha(\omega)\eq\frac12\sum_\pm   \frac{(\omega\pm i   h')}{\sqrt{(\omega\pm i h')^2+\gamma_S^2}},\\
\beta (\omega)\eq-\frac i2\sin\theta\sum_\pm \frac{\pm\gamma_S}{\sqrt{(\omega\pm i h')^2+\gamma_S^2}}., 
\end{eqnarray}
where $\theta= \chi-\chi'$ is the relative angle between the magnetization directions of $F$ and $F'$.

Finally,  the Green function for the effective triplet reservoir solving Eq.~\eqref{dotF} reads
\begin{equation}
g_T= \cosh\vartheta\tau_z+i\sinh\vartheta  \sigma_\chi \tau_\varphi, 
\end{equation} 
with \begin{eqnarray}
\cht(\omega) \eq\frac{\omega+\gamma_F\alpha(\omega)}{\sqrt{[\omega+\gamma_F \alpha(\omega)]^2-\gamma_F^2 \beta^2(\omega)}},\\
\sht(\omega)\eq\frac{\gamma_F\beta(\omega)}{\sqrt{[\omega+\gamma_F \alpha(\omega)]^2-\gamma_F^2 \beta^2(\omega)}}.
\end{eqnarray}
The Green function of the effective triplet reservoir is thus described by a single angle $\vartheta$ which depends, however on all the parameters ($h',\gamma_S,\gamma_F, \theta  $). Note that $\cosh\vartheta$ corresponds to the normal Green function and encodes the density of states, whereas  $\sinh\vartheta$ corresponds to the anomalous Green function, describing the induced triplet correlations. As $\beta\propto\sin\theta$, we see that the triplet correlations vanish for collinear $F'/F$ layers ($\theta=0\,[\pi]$) as expected, while they are maximal for perpendicular magnetizations ($\theta=\pi/2$). For simplicity, we will consider only the case $\theta=\pi/2$ in the following. The generalization to arbitrary angles is straightforward.

Knowing the Green function of the effective triplet reservoir, we can now obtain its density of states (DoS),
\begin{equation}
\nu(\epsilon) = \nu_0 \re[\cht(-i \epsilon+0^+)],
\end{equation}
where $\nu_0$ is the density of states of the normal metal. As the DoS is even in $\epsilon$, we will consider positive energies, $\epsilon>0$, only.

The functions $\alpha (-i\epsilon)$ and $\beta(-i\epsilon)$ possess singularities at $\epsilon=E_c^\pm\equiv |h'\pm\gamma_S|$, which are inherited by the DoS. We will concentrate on the limiting cases $h'\ll\gamma_{S}$ and $h'\gg\gamma_{S}$, when these singularities are far away from $\epsilon=0$. 
In particular, for $\epsilon,\gamma_F,h'\ll \gamma_S$, we find
\begin{equation}
\label{ed:dos1}
\nu(\epsilon)\approx \nu_0 \left[1+ \frac 12\left(\frac {\gamma_Fh'}{\gamma_S^2}\right)^2\left(1+3 \frac {\epsilon^2} {\gamma_S^2}  \right)\right].
\end{equation}
Thus, the zero-energy DoS is enhanced as compared to the normal state. Furthermore, it displays
a broad dip at $\epsilon=0$.
In the opposite regime, for $\epsilon,\gamma_F,\gamma_S\ll h'$, we find
\begin{equation}
\label{ed:dos2}
\nu(\epsilon)\approx \nu_0 \left[1+ \frac 12\left(\frac {\gamma_S}{h'}\right)^2 \frac {1-\frac{\epsilon^2}{\gamma_F^2}}{\left(1+\frac{\epsilon^2}{\gamma_F^2}\right)^2}\right].
\end{equation}
Here as well, the zero-energy DoS is enhanced. However, it possesses a narrow peak at $\epsilon=0$. 
The enhancement of the DoS with respect to its value in the normal state, as well as a peak at $\epsilon=0$, were discussed in similar models of $S/F'/F$ structures with large exchange fields.~\cite{tanaka, braude,fominov2007,Linder2010, kawabata}
 
Similarly, we may analyze the triplet correlations encoded in $\sinh\vartheta(\omega)$. For $\gamma_F,h'\ll \gamma_S$, 
\beq
\sinh\vartheta(\omega) \approx -\frac {\gamma_Fh'}{\gamma_S^2}\frac1{\left(1+\frac{\omega^2}{\gamma_S^2}\right)^{3/2}}.
\label{eq-sh-gamma_S}
\eeq
Thus, the correlations decay on the energy scale $\gamma_S$. By contrast, for $\gamma_F\ll\gamma_S\ll h'$,
\beq
\sinh\vartheta(\omega) \approx -\frac {\gamma_S}{h'} \frac 1{1+\frac\omega{\gamma_F}}\frac1{1+\frac{\omega^2}{h'^2}}.
\label{eq-sh-h'}
\eeq
In that case, the correlations are reduced as soon as $\omega>\gamma_F$ and then decay more rapidly on the energy scale $h'$.


\section{$S_T/S/S_T$ junction\label{section_junc}}
 \label{sec-junction}

\subsection{Current-phase relation}

We are now in a position to study   the  effective $S_T/S/S_T$ junction presented in the Introduction, { see Fig.~\ref{Fig:setup}(b).} Within circuit theory, $S$ is a superconducting node of bare critical temperature $T_c$ and  mean level spacing $\delta$. It is  connected to a left and a right  effective triplet reservoir  ($S_T$)  via connectors of conductances $G_L$ and $G_R$, respectively. Then $g$, $g_L$, and $g_R$ are the Green function in the node, the left reservoir, and the right reservoir  respectively. Here, $g_L$ and $g_R$ are the Green functions derived in the previous section, whereas $g$ will be determined in the following.

For simplicity, here we  consider  the effective odd-frequency triplet reservoirs to be identical by choosing  $\vartheta_L(\omega)=\vartheta_R(\omega)= \vartheta(\omega)$ and   $G=G_L=G_R$. However, the reservoirs may have different superconducting phases $\varphi_{L/R}$ and magnetization axes $\chi_{L/R}$.  
Without loss of generality, 
we choose  $\varphi_{L/R}=\pm\varphi/2$ and $\chi_{L/R}= \pm \chi/2$, such that  $\varphi$ is the phase bias of the junction whereas $\chi$ is the relative angle between the magnetization axes. Then $g_{L/R}$ may be written as
\begin{equation}
g_{L/R}=\cht \tau_z+i\sht  \sigma_{\pm\chi/2}\tau_{\pm\varphi/2}.
\end{equation}
 According to   \refE{dot},  $g$  obeys\begin{equation}\label{eq} 
[\omega\tau_z+ \Delta\tau_\phi + \gamma\frac {g_L+ g_R}2, g]=0 ,
 \end{equation}
where $\gamma= \delta G/(2\pi G_Q)$.  Furthermore, $\Delta$ and $\phi$ are the amplitude and the phase of the order parameter in $S$, satisfying  \refE{self}. 

We concentrate on the weak coupling regime, $\gamma\ll T$, where it is possible to perform a perturbative expansion of $g$ around its bulk value $g_0$. To this end, we write  $ g= g_0+ g_1+\dots$, where $g_1\ll g_0$. Accordingly, the charge current may be written in the form  $I_{L/R}=I_{L/R}^{(1)}+I_{L/R}^{(2)}+\dots$, where
\begin{equation}
\label{currentexp}
I_{L/R}^{(i)}=  \frac{G}{2e}\pi T \, \im \sum_{\omega>0} \frac12\Tr\left[\tau_z[g_{i-1}, g_{L/R}]\right].
\end{equation} 
The  bare Green function of the superconducting node $S$ reads 
 \begin{equation} \label{go}
 g_0=\frac{\omega\tau_z +  \Delta_0\tau_\phi}{\sqrt{\omega^2+\Delta_0^2}},
 \end{equation}
where $\Delta_0(T)$ solves { the standard BCS equation},
\begin{equation}\label{deltao}
 \ln \frac{T_c}  T =2\pi T \sum_{\omega>0} \left(  \frac 1 \omega - \frac1{\sqrt{\omega^2+\Delta_0^2}}\right),
\end{equation}  
while the phase $\phi$   is undetermined for the bare node. However,  the $U(1)$ symmetry is broken once the node is coupled to the reservoirs.

Incorporating $g_0$ in \refE{currentexp}, we obtain  $ I^{(1)}=0$. Namely, no Josephson coupling exists at first order. Indeed, due to the symmetry mismatch between the effective triplet reservoirs and the singlet superconducting dot, a single Cooper pair may not carry a current. 
We thus turn to the next order and compute $g_1$. It obeys the first order expansion of  \refE{eq} supplemented by the normalization condition, namely,
\begin{eqnarray} \label{normo}
[\omega\tau_z+ \Delta_0 \tau_\phi,g_1]&=& -[\gamma\frac {g_L+g_R} 2 + \Delta_1\tau_\phi,g_0],\\\{g_0,g_1\}&=&0. 
 \end{eqnarray} 
 Here $\Delta_1$ is the first order correction to $\Delta_0$.  
The system is solved by 
\begin{eqnarray}
\label{g1}
g_1\eq - \frac 1 {2\sqrt{\omega^2+\Delta_0^2}}\left[\frac\gamma2\left(g_L+g_R\right) +\Delta_1\tau_\phi,g_0\right]g_0.\nonumber
\end{eqnarray} 
Additionally, the self-consistency equation  \eqref{self} yields
\begin{equation}\label{delta1}
\frac{\Delta_1}{\Delta_0}=-  \gamma\left (\sum_\omega\frac{\omega\cht(\omega)}{(\omega^2+\Delta_0^2)^{3/2}}\right)/\left (\sum_\omega \frac {\Delta_0^2}{(\omega^2+\Delta_0^2)^{3/2}}\right).
\end{equation} 
Note that $\Delta_1/\Delta_0<0$, i.e., superconductivity is weakened by  the coupling. This reduction does not depend on the phase bias $\varphi$, and may  be attributed to the gapless property of the odd-frequency triplet  reservoirs that was discussed in the end of Sec.~\ref{sec-reservoirs}, cf. Eqs.~\eqref{ed:dos1} and \eqref{ed:dos2}. It features an inverse proximity effect, where the  quasi-particles, existing  at zero energy in the leads, weaken superconductivity in the node $S$. 
By consequence, the effective critical  temperature $T_c^*$ of $S$ at finite $\gamma$ is decreased, 
\begin{equation}
\frac{T_c^*- T_c}{T_c}\approx-2\pi T\sum_{\omega>0}\frac {\gamma  \cht(\omega)}{\omega^2}<0.
\end{equation}
Note that a dependence of $T_c^*$ on the relative orientation of the magnetizations in adjacent layers would arise in higher order in $\gamma$.

Then, incorporating $ g_1$ in \refE{currentexp}, we find 
\begin{eqnarray}
\label{currentLR}
I^{(2)}\eq-\frac{ \gamma G}{2e}\Big\{a(T) \cos\chi \sin\varphi\\
&&-b(T) \left[\sin\varphi\cos(2\phi)-(\cos\chi+\cos\varphi)\sin{(2\phi)}\right]\Big\},\nonumber
\end{eqnarray}
where 
\begin{eqnarray}
a(T)&= &\pi T \sum_{\omega>0} \frac {  \sinh^2\vartheta(\omega)}{ \sqrt{\omega^2+\Delta_0^2}} \left(2-\frac {\Delta_0^2}{\omega^2+\Delta_0^2}\right),\label{aT}\\
b(T)&=& \pi T\sum _{\omega>0} \frac {   \sinh^2\vartheta(\omega) \,\Delta_0^2}{({\omega^2+\Delta_0^2})^{3/2}} . \label{bT}
\end{eqnarray}
The first line of \eqref{currentLR} may be identified as a quasi-particle current between the two triplet reservoirs that does not depend on the phase of the central island. By contrast, the second line of \eqref{currentLR} may be identified as a condensate contribution, corresponding to a superharmonic Josephson effect between the triplet reservoirs and the singlet central island, that depends on the phase $\phi$.

Current conservation fixes the phase $\phi=k\pi/2$ with $k\in\mathbb{Z}$. Furthermore, energy minimization  imposes
\begin{equation}
\phi=\begin{cases}
0\,\qquad & \mbox{if}\;\cos\varphi +\cos\chi>0,\\ 
\pi/2\,& \mbox{otherwise}.\end{cases}\label{phase}
\end{equation} 
As a consequence of  the odd/even-frequency Josephson coupling between two triplet/singlet pairs,  $\phi$ is defined modulo $\pi$ instead of $2\pi$. 

\begin{figure}[h]
\includegraphics[width=0.9\linewidth]{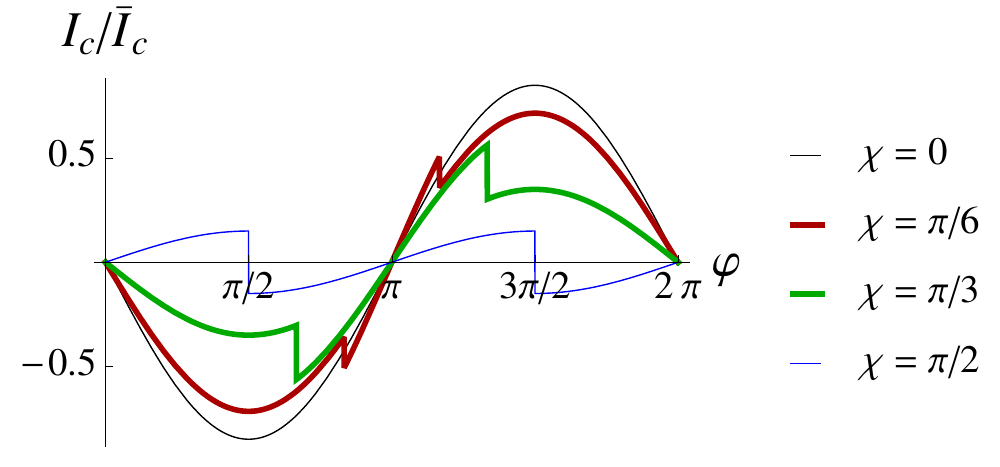}
\caption{Typical current-phase relations. The current is plot in units of $\bar I_c=\gamma G a(T)/(2e)$ for different magnetization angles $\chi$ and $b(T)/a(T)=0.15$, cf.~Eq.~\eqref{eq:current-phase}. The critical current is achieved at phase $\varphi=\pi/2$ for $\chi=0$ and $\pi/6$, and at phase $\varphi=\pi-\chi$ for $\chi=\pi/3$ and $\pi/2$.
\label{fig-cp}}
\end{figure}

Inserting \refE{phase} into \refe{currentLR}, we obtain~\cite{footnote1}
\begin{equation}
I^{(2)}=-{ \frac{ \gamma G}{2e}}\left[a(T) \cos\chi
-b(T){\rm sign}(\cos\chi+\cos\varphi)\right]
\sin\varphi.
\label{eq:current-phase}
\end{equation}
While the quasiparticle contribution is a continuous function of the phase bias $\varphi$, the condensate contribution displays a jump at $\cos\chi+\cos\varphi=0$.~\cite{footnote2} Both $a(T),b(T)\geq0$. Thus, when $\chi<\pi/2$ ($\chi>\pi/2$), the two contributions are opposed for phases $\varphi<\pi-\chi$ ($\varphi<\pi-\chi$) whereas they have the same direction for phases $\varphi>\pi-\chi$ ($\varphi<\pi-\chi$). Examples of typical current-phase relations are shown in Fig.~\ref{fig-cp}.

\subsection{Critical current}
\label{Ic}

To get more insight into the competition between the condensate and quasiparticle contributions, we study the critical current $I_c$ of the junction, where $I_c(T,\chi)=\max_{\varphi}[I^{(2)}(T,\varphi,\chi)]$. Based on the considerations above, the critical current is achieved for phase bias $\varphi=\pi/2$ or $\varphi=\pi-\chi$. Namely, $I_c(T,\chi)=\max[I_{1} , I_{2}]$, where
\begin{eqnarray*}
I_{1}(T,\chi)&=&|I^{(2)}(T,\pi/2,\chi)|={ \frac{ \gamma G}{2e}}\big| a(T) |\cos\chi|-b(T) \big|,\\ 
I_{2}(T,\chi)&=&|I^{(2)}(T,\pi\!-\!\chi,\chi)|\\
&=&{ \frac{ \gamma G}{2e}}\left[
a(T)|\cos\chi |+b(T) \right]\sin\chi.
\end{eqnarray*}
{  For a fixed $\chi$, as a function of temperature, the critical current  $I_c$ lies either on the $I_{1}$ or on the $I_{2}$ branch.  While the $I_2$-branch increases monotonously with decreasing temperature, the temperature dependence of the $I_1$-branch is more complicated. Above $T_c^*$, the $I_1$-branch increases {  as $\ln(\gamma_S/T)$, for $h',T\ll\gamma_S$, and as $(\gamma_F/T)^2$, for $\gamma_F\ll\gamma_S,T\ll h'$}, with decreasing temperature. At $T_c^*$, it has a cusp. When further decreasing temperature below $T_c^*$, it increases much more slowly or even decreases. For $\gamma_S\gg h'$ and $T\lesssim T_c^*$, we find 
\begin{eqnarray}
\label{eq:I1a}
I_{1}(T,\chi)&\simeq&{ \frac{ \gamma G}{2e}}\left(\frac{\gamma_Fh'}{\gamma_S^2}\right)^2\Big\{|\cos\chi|\ln\frac{\gamma_S}T\\
&&\qquad\qquad-\frac{(T_c^*-T)}{T_c^*}(2|\cos\chi|+1)\Big\},\nonumber
\end{eqnarray}
which decreases with decreasing temperature for all values of $\chi$.
By contrast, for $\gamma_S\ll h'$  and $T\lesssim T_c^*$, we find
\begin{eqnarray}
\label{eq:I1b}
I_{1}(T,\chi)&\simeq&{ \frac{ 7\gamma G}{8\pi^2e}}\zeta(3)\left(\frac{\gamma_S}{h'}\right)^2\left(\frac{\gamma_F}{T}\right)^2\Big\{|\cos\chi|\\
&&\qquad\qquad-{\cal N}\frac{(T_c^*-T)}{T_c^*}(2|\cos\chi|+1)\Big\},\nonumber
\end{eqnarray}where ${\cal N}=31\zeta(5)/[7\zeta(3)]^2\approx0.5$, which slowly increases with decreasing temperature for angles $\chi \lesssim \pi/3$.

Which branch the critical current follows is determined by the ratio $b(T)/a(T)$. We find that the ratio $b(T)/a(T)$ is zero above $T_c^*$ and  increases monotonously below $T_c^*$ , satisfying  $b(T)/a(T)<1$.

As a consequence, at high temperatures, the critical current follows the $I_1$-branch. 
At lower temperatures, one  may distinguish two different behaviors depending on whether
$I_1(0,\chi)$ is larger or smaller than $I_2(0,\chi)$.
 The critical angle $\chi_c$ at which one switches between the two cases is given by
\begin{equation}
  \frac {b(0)}{a(0)}=  \frac {|\cos\chi_c|(1-\sin\chi_c)}{1+\sin   \chi_c}.
  \label{eq:criterion-chic}
  \end{equation} 
The solution $\chi_c$ of this equation increases from 0 to $\pi/2$ as $b(0)/a(0)$ decreases from { 1 to 0}, cf.~the dependence of the r.h.s. of \eqref{eq:criterion-chic} as a function of $\chi_c$ in Fig.~\ref{Fig:chi}.
For angles $\chi<\chi_c$, the critical current lies on the $I_1$-branch at all temperatures. 
By contrast, for angles $\chi>\chi_c$, the current switches to the $I_2$-branch at the temperature $T_{12}$ determined by 
\begin{equation}
  \frac {b(T_{12})}{a(T_{12})}=  \frac {|\cos\chi|(1-\sin\chi)}{1+\sin   \chi}.
  \end{equation} 
Using these considerations, we now consider the temperature dependence of the critical current for the cases $h'\ll\gamma_S$ and $h'\gg\gamma_S$, assuming $\gamma_F\ll T_c^*\ll\max[\gamma_S,h']$. 

\begin{figure}[h]
\includegraphics[width=0.9\linewidth]{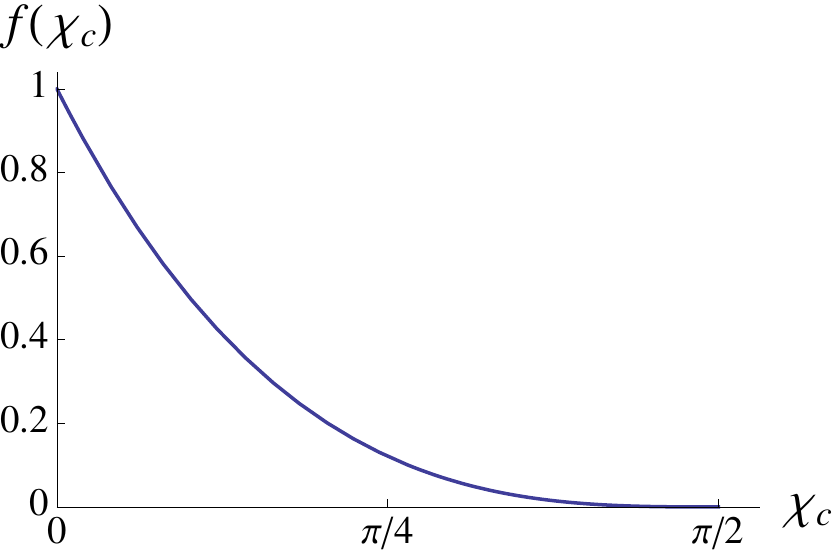}
\caption{ Plot of the r.h.s.~of \eqref{eq:criterion-chic}, $f(\chi_c)= |\cos\chi_c|(1-\sin\chi_c)/(1+\sin   \chi_c)$, as a function of $\chi_c$.
}
\label{Fig:chi}
\end{figure}

For $h'\ll\gamma_S$, $b(0)/a(0)$ decreases { as $1/\ln(\gamma_S/T_c^*)$} with decreasing $T_c^*/\gamma_S$. Thus, the critical angle $\chi_c$ increases. Since $I_1$ decreases with decreasing temperature below $T_c^*$, the critical current displays a unique maximum at temperatures close to $T_c^*$ for a wide range of angles. This non-monotonous temperature dependence provides a clear signature of the competition between odd/odd- and odd/even-frequency couplings.

\begin{figure}[h]
\includegraphics[width=0.9\linewidth]{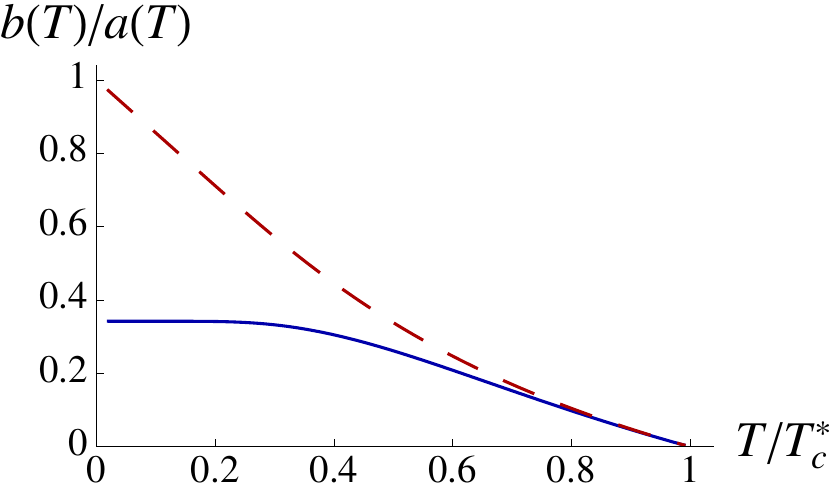}
\caption{ Plot of $b(T)/a(T)$ as a function of $T/T_c^*$, for $h',T_c^*\ll\gamma_S$ (solid line) and $\gamma_S,T_c^*\ll h'$ (dashed line).}
\label{Fig:ratio_ba}
\end{figure}

By contrast, for $\gamma_S\ll h'$, we obtain $b(0)/a(0)\approx1$. Thus, a finite temperature $T_{12}$ below which the critical current starts rising again rapidly exists for all angles. In the intermediate temperature regime $T_{12}<T<T_c^*$, the current increases very slowly. Though less pronounced than in the opposite parameter regime, this peculiar temperature dependence is a signature of the competition between the different symmetry couplings.}

\begin{figure}[h]
\includegraphics[width=0.9\linewidth]{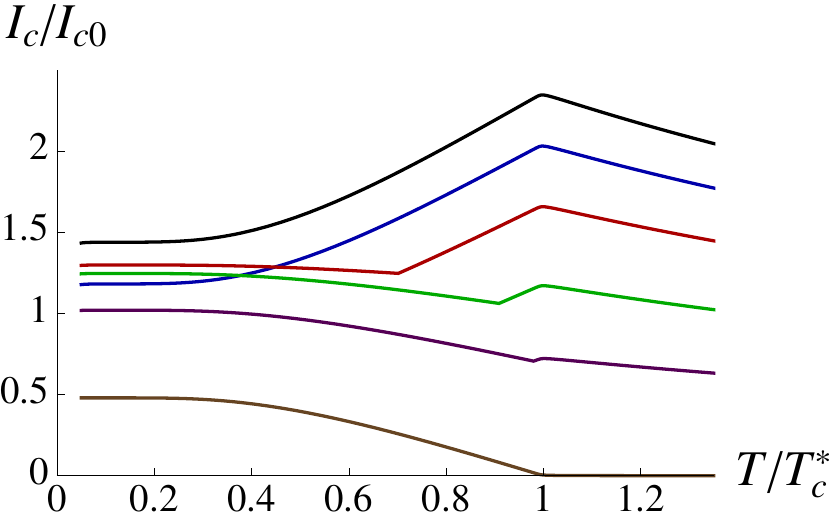}
\caption{ Plot of the critical current $I_c$ [in units of $I_{c0}=\gamma G (\gamma_F h'/\gamma_S^2)^2/(2e)$] as a function of $T/T_c^*$, for $h',T_c^*\ll\gamma_S$, and different angles $\chi=0,\pi/6,\pi/4,\pi/3,2\pi/5,\pi/2$ (from top to bottom).}
\label{Fig:large_gamma}
\end{figure}

\begin{figure}[h]
\includegraphics[width=0.9\linewidth]{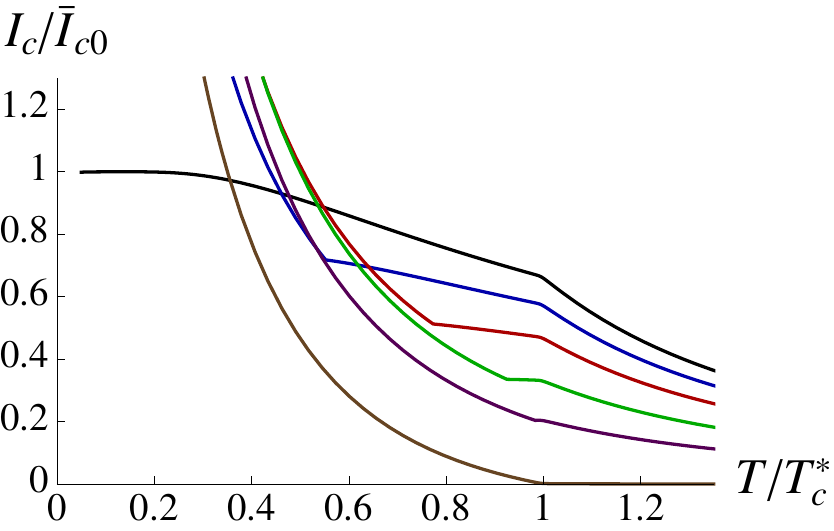}
\caption{ Plot of the critical current $I_c$ [in units of $\bar I_{c0}=\gamma G [\gamma_F \gamma_S/( h' \Delta_0^*)]^2/(2e)$], with $\Delta_0^*\approx 1.76 T_c^*$, as a function of $T/T_c^*$, for $\gamma_F\ll \gamma_S,T_c^*\ll h'$, and different angles $\chi=0,\pi/6,\pi/4,\pi/3,2\pi/5,\pi/2$ (from top to bottom). Note that the critical current saturates at $T\lesssim \gamma_F$ (not visible on the scale of the figure).}
\label{Fig:large_h}
\end{figure}

{ The temperature dependence of the critical current described above is illustrated in Figs.~\ref{Fig:large_gamma} and  \ref{Fig:large_h}, for the cases $h'\ll\gamma_S$ and $h'\gg\gamma_S$, respectively, and for different angles $\chi$.}

\section{Metallic junction}
\label{sec-metallic}

While our calculations are limited to tunnel contacts, we may generalize our considerations to metallic junctions based on the identification of the different contributions to the current. To do so, we examine the corresponding terms in the Josephson energy. The current is then obtained by taking a derivative with respect to phase. As before, we assume that the angle between the magnetizations of adjacent ferromagnets is $\theta=\pi/2$, cf.~\onlinecite{footnote1}.

For a metallic $S'/F'/F/S/F/F'/S'$ junction, the quasiparticle contribution near the critical temperature of the $S$ layer takes the form
\begin{equation}
E_J^{\rm qp}=g\left(\Delta'-c^{\rm qp} \frac{\Delta^2(T)}{T_c^*}\right)e^{-L_S/\xi_S}\cos\chi\cos(\varphi_L-\varphi_R),
\label{eq:metallic-qp}
\end{equation}
where $g\sim G/G_Q$, $G$ is the normal-state conductance of the junction, and $c^{\rm qp}$ is a numerical factor of the order of unity. Here we write the phases $\varphi_L$ and $\varphi_R$ of the left and right superconductors $S'$ explicitly. Furthermore, $L_S$ and $\xi_S\sim\sqrt{D/ T_c^*}$ are the length and coherence length of the central superconductor, respectively, where $D$ is the diffusion coefficient. The first term $\propto \Delta'$ describes the usual Josephson energy when the central superconductor is in the normal state. It increases monotonously as the temperature decreases below the critical temperature $T_c'$ of the leads and saturates at temperatures $T\ll T_c'$. Thus, close to $T_c^*\ll T_c'$, we may neglect its temperature dependence.~\cite{footnote3}The second term $\propto\Delta^2(T)/T_c^*$ accounts for the reduction the quasiparticle contribution (responsible for the {\it triplet} supercurrent flow) due to the developing of {\it singlet} superconducting correlations below $T_c^*$, when $\Delta(T)\propto \sqrt{T_c^*-T}\,\theta(T_c^*-T)$ is finite.

The condensate contribution consists of three different terms. Namely, there is a Josephson coupling between the left superconductor and the central superconductor, depending on the phase difference $\varphi_L-\phi$, as well as a Josephson coupling between the central superconductor and the right superconductor, depending on the phase difference $\phi-\varphi_R$.  As the first harmonic in a bilayer junction is short-ranged, only the second harmonic survives for both of these contributions. Furthermore, there is a crossed term, where two pairs from each of the outer superconductors recombine in the central superconductor. Thus this contribution depends on the phase $\varphi_L+\varphi_R-2\phi$. It is suppressed with the length of the central superconductor on the scale of the coherence length, and it depends on the angle between the magnetizations of the left and right ferromagnet. As a consequence, the condensate contribution takes the form
\begin{eqnarray}
E_J^{\rm cond}&=&-c^{\rm cond} g\frac{\Delta^2(T)}{T_c^*}\Big\{\cos\left[2(\varphi_L-\phi)\right]
\\
&&+\cos\left[2(\phi-\varphi_R)\right]
\nonumber\\
&&+2e^{-L_S/\xi_S}\cos\chi\cos(\varphi_L+\varphi_R-2\phi)\Big\},
\nonumber
\label{eq:metallic-cd}
\end{eqnarray}
where $c^{\rm cond}\propto \xi_F/L$.~\cite{superh} With $\varphi_L=-\varphi_R=\varphi/2$, the expression simplifies to
\begin{eqnarray}
E_J^{\rm cond}&=&-2 c^{\rm cond}g\frac{\Delta^2(T)}{T_c^*}\Big\{\cos\varphi+e^{-\frac{L_S}{\xi_S}}\cos\chi\Big\}\cos(2\phi).
\nonumber\\
\end{eqnarray}
Then, minimization of the energy with respect to $\phi$ yields
\begin{equation}
\phi=\begin{cases}
0\,\qquad & \mbox{if}\, \cos\varphi +e^{-L_S/\xi_S}\cos\chi>0,\\ 
\pi/2\,& \mbox{otherwise}.\end{cases}\label{phase2}
\end{equation} 

Finally, the supercurrent accounting for both the quasiparticle and condensate contributions reads
\begin{eqnarray}
\label{eq:Imetal}
I&=&-G\left\{\left[\Delta'-c^{\rm qp}\frac{\Delta^2(T)}{T_c^*}\right]e^{-L_S/\xi_S}\cos\chi\right.
\nonumber \\
&&-\left.2 c^{\rm cond}\frac{\Delta^2(T)}{T_c^*}{\rm sign}\left(\cos\varphi +e^{-L_S/\xi_S}\cos\chi\right)\right\}\sin\varphi.
\nonumber \\
\end{eqnarray}
Eq.~\eqref{eq:Imetal} has a similar form as the current-phase relation Eq.~\eqref{eq:current-phase} in the tunneling regime. For a short $S$ layer with $L_S\ll \xi_S$, we find that the critical current is given by $I(\varphi=\pi/2)$, corresponding to the $I_1$-branch discussed in Sec.~\ref{Ic}. Thus, the critical current decreases below $T_c^*$ as
\begin{equation}
I_c=G\left\{\Delta'|\cos\chi|-\frac{\Delta^2(T)}{T_c^*}\Big[c^{\rm qp} |\cos\chi|+2c^{\rm cond} \Big]\right\}
\end{equation}
(except for angles $\chi\sim\pi/2$). As in the tunneling regime, this peculiar temperature dependence provides a clear signature of the competition between odd/odd- and odd/even-frequency couplings.
By contrast, in the opposite regime, $L_S\gg\xi_S$, the temperature regime below $T_c^*$, where the $I_1$-branch dominates, shrinks to zero. Thus,   
the critical current is given by $I(\varphi=\pi/2+0^+)$, corresponding to the $I_2$-branch discussed in Sec.~\ref{Ic}. We obtain
\begin{equation}
I_c=G\left\{\Delta'|\cos\chi|e^{-L_S/\xi_S}+2c^{\rm cond} \frac{\Delta^2(T)}{T_c^*}\right\}.
\end{equation}
Here, the quasiparticle and condensate contributions add up, which leads to an enhancement of $I_c$ below $T_c^*$.

\section{conclusion}
\label{sec-conclusion}

Using circuit theory, we have proposed a simple model for the Green function $g_T$ of an effective triplet odd-frequency superconducting reservoir $S_T.$   Then, we have studied the coexistence of singlet even-frequency and triplet odd-frequency superconducting correlations in an $S_T/S/S_T$ Josephson junction. 

{  We predict that the competition between odd/odd-frequency  and odd/even-frequency Josephson couplings may be observed in a peculiar temperature dependence of the critical current of the $S_T/S/S_T$ junction below the transition temperature $T_c^*$ of the central superconductor. For a large range of parameters, the critical current either increases very slowly or even decreases when lowering the temperature below $T_c^*$.}  This is in sharp contrast with a conventional $S'/S/S'$ junction, where the superconducting transition of the central superconductor leads to an enhancement of the critical current.~\cite{kuprianov}

We propose to realize such an $S_T/S/S_T$ junction by fabricating a hybrid $S'/F'/F/S/F/F'/S'$ junction, i.e., by inserting a superconducting layer in the middle of an  $S/F'/F/F'/S$ junction such as the ones presented in Refs.~\onlinecite{robinson, birge}.  

\acknowledgements

We thank Jason Robinson for helpful discussions on related experiments. Furthermore, we acknowledge support through ANR Grants No.~ANR-11-JS04-003-01 and No.~ANR-12-BS04-0016-03, the NanoSC COST Action MP1201, and an EU-FP7 Marie Curie IRG.
 
 \appendix

\section{Derivation of the Green function in the strong ferromagnet $F$}
\label{app-hinf}

At large $h$, the solution of the equation 
\begin{equation}
\label{eqbigh}
[h\sigma_z\tau_z + A,g]=0,
\end{equation}
with $g^2=1$, can be expanded perturbatively in $1/h$: 
$g=g^{(0)}+(1/h) g^{(1)}+\dots$
Let us now introduce the decomposition $A=A_\perp+A_\parallel$, with 
$[A_\parallel,\sigma_z\tau_z]=0$ and $\{ A_\perp,\sigma_z\tau_z\}=0$. Similarly, $g^{(n)}=g^{(n)}_\parallel+g^{(n)}_\perp$.

In the leading order in $h$, Eq.~\eqref{eqbigh} yields $g^{(0)}_\perp=0$, while the normalization condition reads $(g_\parallel^{(0)})^2=1$.

In the next order, Eq.~\eqref{eqbigh} yields
\begin{equation}
2\sigma_z\tau_zg^{(1)}_\perp+[A,g^{(0)}_\parallel]=0,
\end{equation}
while the normalization condition reads $\{g_\perp^{(1)}+g_\parallel^{(1)},g_\parallel^{(0)}\}=0$.
It is solved with $g_\parallel^{(0)}= A_\parallel/\sqrt{A_\parallel^2}$ (note that $A_\parallel^2$ is scalar), 
$g_\perp^{(1)}=-(1/2) \sigma_z\tau_z[A_\perp, g_\parallel^{(0)} ]$, and $g_\parallel^{(1)}=0$.


\begin{thebibliography}{99}
\bibitem{footnote1} For an arbitrary angle $\theta$ between $F$ and $F'$ layers in the triplet reservoirs, the current \eqref{eq:current-phase} is reduced by an additional factor $\sin^2\theta$.
\bibitem{footnote2} 
Note that in an asymmetric junction, $\gamma_L\neq\gamma_R$, the jump is rounded on a scale $\delta\varphi\sim (\gamma_L-\gamma_R)/(\gamma_L+\gamma_R)$.
\bibitem{footnote3} {This is in contrast with the tunneling regime where the temperature dependence of the corresponding term is determined by the smaller scale $\gamma$ and, thus, needed to be calculated explicitly near $T_c^*$, cf.~the  first term in Eqs.~\eqref{eq:I1a} or \eqref{eq:I1b}.}
\end{thebibliography}
\end{document}